\documentclass[%
preprint,
 amsmath,amssymb,
 aps,
prx,
]{revtex4-2}

\usepackage{graphicx}
\usepackage{bm}
\usepackage{hyperref}
\usepackage{upgreek}
\usepackage[mathlines]{lineno}

\begin{document}


\title{Visualization of Topological Boundary Modes Manifesting Topological Nodal-Point Superconductivity}



\author{Abhay Kumar Nayak}
\thanks{These three authors contributed equally}
\affiliation{Department of Condensed Matter Physics, Weizmann Institute of Science, Rehovot 7610001, Israel.}

\author{Aviram Steinbok}
\thanks{These three authors contributed equally}
\affiliation{Department of Condensed Matter Physics, Weizmann Institute of Science, Rehovot 7610001, Israel.}

\author{Yotam Roet}
\thanks{These three authors contributed equally}
\affiliation{Department of Condensed Matter Physics, Weizmann Institute of Science, Rehovot 7610001, Israel.}

\author{Jahyun Koo}
\affiliation{Department of Condensed Matter Physics, Weizmann Institute of Science, Rehovot 7610001, Israel.}

\author{Gilad Margalit}
\affiliation{Department of Condensed Matter Physics, Weizmann Institute of Science, Rehovot 7610001, Israel.}

\author{Irena Feldman}
\affiliation{Department of Physics, Technion - Israel Institute of Technology, Haifa 32000, Israel.}

\author{Avior Almoalem}
\affiliation{Department of Physics, Technion - Israel Institute of Technology, Haifa 32000, Israel.}

\author{Amit Kanigel}
\affiliation{Department of Physics, Technion - Israel Institute of Technology, Haifa 32000, Israel.}

\author{Gregory A. Fiete}
\affiliation{Department of Physics, Northeastern University, Boston, Massachusetts 02115, USA.}
\altaffiliation{Department of Physics, Massachusetts Institute of Technology, Cambridge, Massachusetts 02139, USA.}

\author{Binghai Yan}
\affiliation{Department of Condensed Matter Physics, Weizmann Institute of Science, Rehovot 7610001, Israel.}

\author{Yuval Oreg}
\affiliation{Department of Condensed Matter Physics, Weizmann Institute of Science, Rehovot 7610001, Israel.}

\author{Nurit Avraham}
\email[]{nurit.avraham@weizmann.ac.il}
\affiliation{Department of Condensed Matter Physics, Weizmann Institute of Science, Rehovot 7610001, Israel.}

\author{Haim Beidenkopf}
\email[]{haim.beidenkopf@weizmann.ac.il}
\affiliation{Department of Condensed Matter Physics, Weizmann Institute of Science, Rehovot 7610001, Israel.}


\date{\today}

\begin{abstract}
The extension of the topological classification of band insulators to topological semimetals gave way to the topology classes of Dirac, Weyl and nodal line semimetals with their unique Fermi arc and drum head boundary modes \cite{Hasan2010,Chiu2016,Armitage2018}. Similarly, there are several suggestions to employ the classification of topological superconductors for topological nodal superconductors with Majorana boundary modes \cite{Qi2011,Schnyder2015,Sato2017}. Here, we show that the surface 1H termination of the transition metal dichalcogenide compound 4Hb-TaS$_2$, in which 1T-TaS$_2$ and 1H-TaS$_2$ layers are interleaved, has the phenomenology of a topological nodal point superconductor. We find in scanning tunneling spectroscopy a residual density of states within the superconducting gap. An exponentially decaying bound mode is imaged within the superconducting gap along the boundaries of the exposed 1H layer characteristic of a gapless Majorana edge mode. The anisotropic nature of the localization length of the edge mode aims towards topological nodal superconductivity. A zero-bias conductance peak is further imaged within fairly isotropic vortex cores. All our observations are accommodated by a theoretical model of a two-dimensional nodal Weyl-like superconducting state, which ensues from inter-orbital Cooper pairing. The observation of an intrinsic topological nodal superconductivity in a layered material will pave the way for further studies of Majorana edge modes and its applications in quantum information processing \cite{Kitaev2003,Nayak2008,Stern2013,Lian2018}. 
\end{abstract}


\maketitle

\section{main}

Topological superconductors are extensively explored with the aim to induce and manipulate Majorana zero modes that are essential to realize topological quantum information processing. The nonlocal nature of these zero modes and their intrinsic non-Abelian braiding statistics allow for quantum information to be stored and manipulated in a nonlocal manner, which protects it from the influence of the local environment \cite{Kitaev2003,Nayak2008,Stern2013}. A growing body of realizations of topological superconductivity in one-dimensional (1D) systems, including hybrid nanowires \cite{Mourik2012,Das2012,Lutchyn2017,Vaitiekenas2020}, atomic chains \cite{Nadj-Perge2014,Kim2018}, proximitized helical edge modes \cite{Jack2019} and planar Josephson junctions \cite{Fornieri2019,Ren2019} have been reported so far. In these systems, Majorana zero modes manifest themselves as zero-energy conductance anomalies localized at the ends of the topological superconducting segment. In 2D systems, Majorana zero modes may appear as zero bias conductance peak at the center of vortex cores \cite{Guan2016,Yin2015,Chen2018a,Wang2018d,Machida2019,Kong2019,Zhu2020,Liu2018,Chen2020a,Liu2020,Yuan2019}. Yet, an unambiguous hallmark of 2D topological superconductors is the existence of gapless Majorana edge modes bound to its 1D boundaries. \cite{Qi2011,Schnyder2015,Sato2017}. These could be the physical boundaries of the system or such that separate superconducting regions with distinct topological nature. So far, signatures of 1D Majorana edge modes have been observed mostly in heterostructures or hybrid systems, \cite{Menard2017,Palacio-Morales2019,Kezilebieke2020, Wang2020}, in which topological superconductivity emerges from the interplay of magnetism, superconductivity and spin-orbit coupling that are provided by the different material components. To date, evidence for topological superconductivity in naturally occurring materials have been hardly reported \cite{Yuan2019,Jiao2020}.

Transition metal dichalcogenides (TMDs) are strongly correlated materials exhibiting numerous different interesting phases such as superconducting, charge density wave (CDW) and spin liquid \cite{Ye2012,Yuan2014,Lu2015,Xi2016,Hsu2017,DelaBarrera2018,Sajadi2018,Fatemi2018,Wickramaratne2020,Devarakonda2020}. The combination of strong spin orbit and strong electron interactions renders it an ideal hunting ground for topological superconductivity \cite{Yuan2014,Zhou2016,Hsu2017,He2018TSC, Fischer2018,Hsu2020,Kanasugi2020,Shaffer2020}. Their van der Waals nature further allows one to mix and match various materials with diverse phenomenology either in growth or via exfoliation and stacking \cite{Geim2013}. We use scanning tunneling microscopy (STM) and spectroscopy to study the superconducting state in the TMD 4Hb-TaS$_2$, which consists of alternating stacking of 1T- and 1H-TaS$_2$ layers (Fig.\ref{fig1}a). While bulk 1T-TaS$_2$ is expected to host a quantum spin liquid state \cite{Law2017}, bulk 2H-TaS$_2$ is well known for its quasi-2D Ising superconductivity \cite{DelaBarrera2018}. Furthermore, the observation of increased muon spin rotation rising concurrently with the superconducting state of 4Hb-TaS$_2$ has raised the possibility of non-trivial topology \cite{Ribak2019}. Our measurements reveal the presence of 1D boundary edge modes localized at crystallographic edges and separate between superconducting regions of distinct topological nature. Their anisotropic nature together with residual in-gap bulk density of states, elude to a nodal-point topological superconducting state that we attribute to inter-orbital pairing mechanism promoted by strong interactions. Zero bias conductance peaks at vortex cores are also detected. Our results are suitably captured within a theoretical model of topological nodal superconductivity stemming from inter-orbital Cooper pairing \cite{Gao2010,Wang2012,Fukaya2018,Schnyder2015}. These observations identifying 4Hb-TaS$_2$ in particular, and complex TMD compounds in general, as a fascinating bedrock for exotic topological superconducting states. 

\section{Results}
\begin{figure} 
\includegraphics[width=1\linewidth]{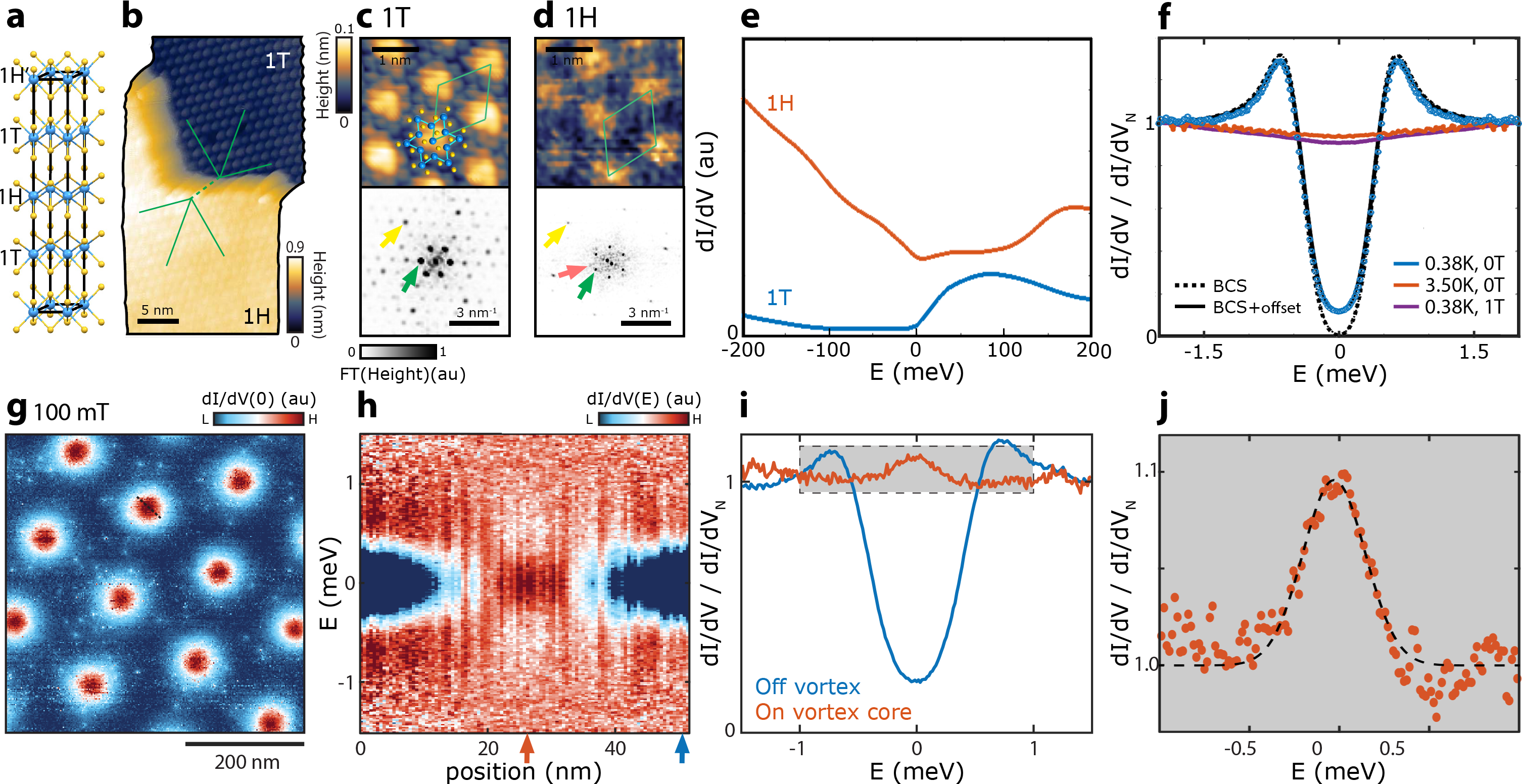} 
\centering
\caption{}  \label{fig1}   
\end{figure}

We cleave single crystals along the [001] orientation under ultra-high vacuum. Due to the alternating layer structure of 4Hb-TaS$_2$ the cleave exposes sulfur surfaces of both 1T and 1H terminations, as well as step edges among them \cite{DiSalvo1973,Tanaka1988,Kim1995}, as shown in Fig.\ref{fig1}b. The two layer terminations can be distinguished by their characteristic charge density wave (CDW) patterns: The 1T layers host a strong $\sqrt{13}\times\sqrt{13}$ CDW reconstruction pattern in which 13 tantalum atoms bunch in a star of David formation that is arranged on a triangular super-lattice, as shown in the topographic image in Fig.\ref{fig1}b and c. The 1H layers host a weak intrinsic $3\times3$ CDW superlattice, superimposed with a strong imprint of the $\sqrt{13}\times\sqrt{13}$ CDW superlattice of the adjacent 1T layers (Fig.\ref{fig1}d) \cite{Fujisawa2018}. The different types of CDWs are distinguished by their corresponding Bragg peaks in the Fourier transform of the topographic images shown in the bottom panels of Fig.\ref{fig1}c and d. The green lines marked across the step edge in Fig.\ref{fig1}b show that the native 1T CDWs on the 1T layer and their imprint on the 1H layer are indeed fully commensurate. The 1T and 1H terminations differ also in their surface density of states (DOS), measured in differential conductance ($dI/dV$), as shown in Fig.\ref{fig1}e. While the 1T spectrum shows low DOS around the Fermi energy and at negative biases, the 1H spectrum is metallic.

We start by exploring the superconducting state that forms at low temperatures in the metallic 1H layer. Spectra measured on the 1H layer at 4.2 K and at a low temperature of 380 mK are shown in Fig.\ref{fig1}f. A clear superconducting gap structure with accompanying coherence peaks is observed at 380 mK (circles). However, the DOS within the gap does not vanish completely but rather exhibits a soft gap with a residual in-gap conductance. This residual DOS cannot be fitted with instrumental and thermal broadening of a Bardeen-Cooper-Schrieffer (BCS) spectrum (dotted line). A finite DOS offset has to be added in order to obtain a tight fit (solid line, See also Fig.S2), yielding a gap of $\Delta = 0.44$ meV. The observed finite in-gap conductance may relate to the residual heat capacity that was previously detected within the superconducting transition \cite{Ribak2019} and its possible origin will be discussed below. The superconducting gap vanishes at temperatures above 2.7 K and at out of plane magnetic fields higher than 0.9 T (Fig.S1e and f, respectively, and Fig.\ref{fig1}f), in agreement with previously reported transport results \cite{Ribak2019}. At magnetic fields lower than the critical one (see 100 mT map in Fig.\ref{fig1}g and 200 mT map in Fig.S5) we image normal vortex cores in zero-bias $dI/dV$ maps. The vortex cores appear fairly isotropic and the DOS decays exponentially over a coherence length of $\xi\approx21$ nm (Fig.S4a), in agreement with values estimated from $H_{c2}\approx\Phi_o\pi\xi^2$, where $\phi_o=h/2e$ is a flux quantum. The vortices form a regular Abrikosov vortex lattice, signifying weak pinning. Intriguingly, in-between vortex cores we find the presence of speckles, where the ZBC is enhanced by $\sim10\%$ (supplementary Fig.S3). At large enough voltage biases to probe the normal state, the speckles have no spectroscopic trace in the DOS (Fig.S3d). Both the residual in-gap DOS as well as its susceptibility to disorder are characteristic of nodal superconductors rather than $s$-wave ones \cite{Pan2000,Balatsky2006,Liu2008,Liu2019,Andersen2020}.

Careful inspection of the $dI/dV$ spectra at vortex cores finds a ZBC peak localized at their centers, as shown in Fig.\ref{fig1}h (see also Fig.S4 and Fig.S5). Representative $dI/dV$ spectra measured on the vortex core and away from it are given in Fig.\ref{fig1}i, along with a magnified image of the ZBC peak at the vortex center (Fig.\ref{fig1}j). A Gaussian fit finds an energy width of $\sigma=0.25$ meV, which corresponds to a temperature scale of about 1.7 K, higher than the electronic temperature in our system which is less than 1 K \cite{Reiner2020} (see also Fig.S2). Recently, similar ZBC peaks at vortex cores have been identified in a few other electronic systems and have been attributed to topological Majorana states \cite{Chen2018a,Wang2018d,Zhu2020,Yuan2019}. In TaS$_2$, however, the small energy level spacing between Caroli-de Gennes-Matricon core states ($\delta\sim\Delta^2/E_F$), set by the high Fermi energy, does not allow one to make a distinction between them and Majorana zero energy states \cite{Hess1989,Gygi1991}. Therefore, while observation of ZBC peaks at vortex cores may indicate topological superconductivity, it alone does not provide a sufficient evidence for the non-trivial topological nature \cite{Pan2000,Wang2004,Nagai2014}.

\begin{figure}
\includegraphics[scale=1.5]{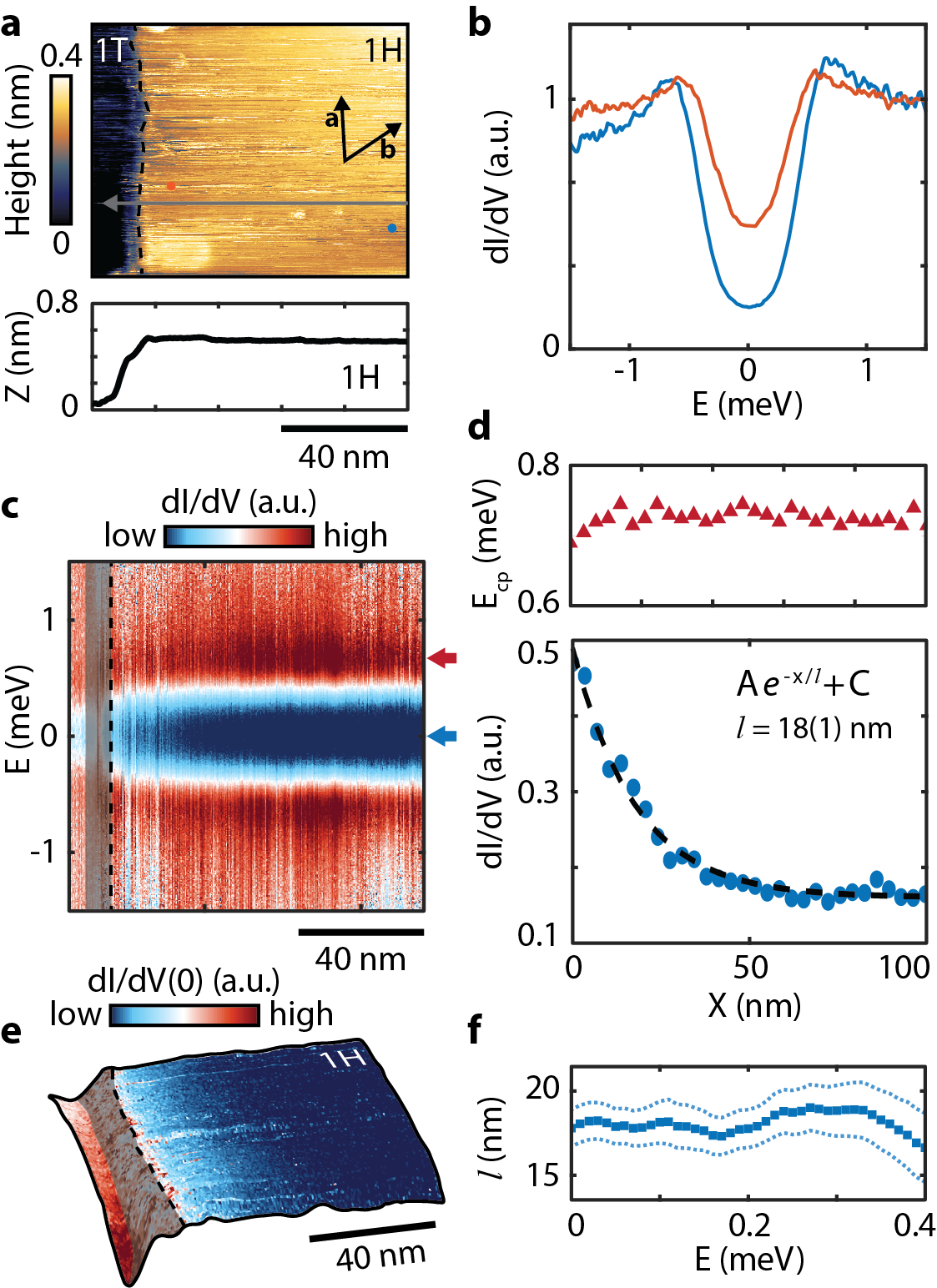} 
\centering
\caption{}   \label{fig2}  
\end{figure}

To further investigate the topological nature of the 1H layer we explored the spectroscopic properties across the layer boundaries provided by both 1H and 1T step edge terminations. A 2D topological superconductor is expected to carry 1D Majorana modes on its boundaries \cite{Hasan2010,Qi2011}. We first examine the step-edge terminated 1H layer shown in Fig.\ref{fig2}a. The $dI/dV$ spectra measured at the step-edge boundary and far from it are shown in Fig.\ref{fig2}b. Far from the step edge (blue line) we find a soft superconducting gap, similar to the one observed on open 1H terraces. In contrast, the spectra measured close to the step edge (red lines) show a much shallower gap. The continuous evolution of the spectra, on approaching the step-edge, is shown in Fig.\ref{fig2}c. The size of the gap, as captured by the coherence peaks ($E_{cp}$), is hardly changed down to the last few atoms away from the step edge while the in-gap DOS increases significantly on approaching the step edge (top and bottom panels of Fig.\ref{fig2}d, respectively). This behaviour indicates that the increased in-gap DOS does not reflect suppression of superconductivity by the step-edge but rather excess in-gap DOS added by the existence of an edge mode there.

The observed boundary mode runs all along the rather irregular step edge, as mapped in Fig.\ref{fig2}e. Its spatial distribution, as extracted from the increase in ZBC on approaching the step edge, fits an exponentially localized profile, $Ae^{-x/l}+C$, with a localization length of $l\approx 18$ nm (dashed line in Fig.\ref{fig2}d). This value is comparable to the coherence length, $\xi\approx 21$ nm, which we extract from the vortex core profiles (Fig.S4a). This exponential localization of the edge mode is not unique to zero bias, but rather remains fairly constant across the superconducting gap as shown in Fig.\ref{fig2}f (see also Fig.S6). This signifies that the edge mode disperses in energy across the superconducting gap, in sharp contrast to the ZBC state at vortex cores that exhibits a peaked energy profile. From the energy independent localization length we estimate the Fermi velocity via $v_F\approx l\Delta/h=2 \times 10^4$ m/sec, where $h$ is Planck's constant and $\Delta$ is the superconducting gap extracted from our measurements. While this velocity is an order of magnitude slower than the bare electron velocity in 1H-TaS$_2$, it is comparable to the renormalized velocity, $v_F\approx 6 \times 10^4$ m/sec, reported previously in the vicinity of the Fermi energy due to electron-phonon interactions \cite{Wijayaratne2017}. When superconductivity is suppressed, either by raising the temperature or by the application of a magnetic field above their respective critical values, we find no traces of a localized mode or any other irregularity in the local DOS near the step edge (see Fig.S7). This shows that the origin of the edge mode lies in the nature of the superconducting states in the exposed 1H-TaS$_2$ layer.

\begin{figure}
\includegraphics[scale=1]{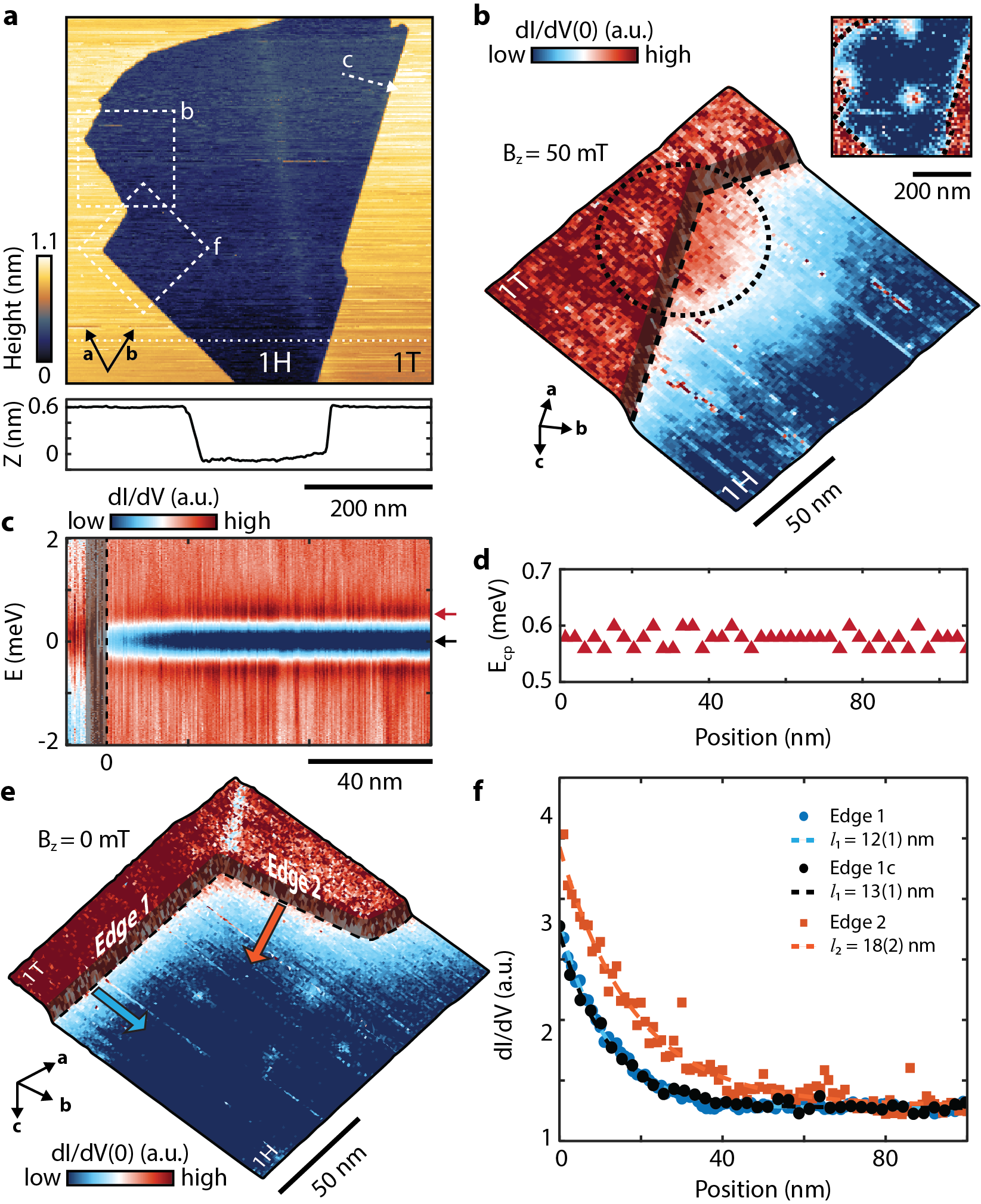} 
\centering
\caption{}     \label{fig3}
\end{figure}

The observation of a dispersing gapless edge mode localized at the step edge terminations of the 1H-TaS$_2$ layer provides a strong indication for the topological nature of the superconducting state. We now show that this topological nature is crucially influenced by the proximity of the 1H-TaS$_2$ layer to the 1T-TaS$_2$ layer, in the 4Hb-TaS$_2$ layer structure. To that end we investigate the boundary between exposed and 1T-encapsulated 1H layers. Such a boundary is displayed in the topographic structure shown in Fig.\ref{fig3}a, in which a crater amongst the 1T layer exposes the 1H layer underneath it. In the crater region the 1H layer is adjacent to a 1T layer only from below, whereas in the crater surroundings it is encapsulated by 1T layers from both above and below. Since 4Hb-TaS$_2$ exhibits bulk superconductivity \cite{Ribak2019} we expect the 1H layer to remain superconducting throughout, yet the top 1T layer is metallic, which will be discussed elsewhere. Indeed, we commonly image portions of vortex cores peeking beyond the top 1T layer termination, as demonstrated in Fig.\ref{fig3}b, signifying 1H-TaS$_2$ superconductivity continues well into the 1T-encapsulated part. Nevertheless, in the face of the continuous superconducting layer, when we map the ZBC at the boundary of the 1H crater we clearly image the presence of an edge mode running along it, as shown in Fig.\ref{fig3}c (see also Fig.S8 in SI). Again, the increased in-gap DOS is not accompanied by a significant change in the superconducting gap size, as extracted from the coherence peaks' energy ($E_{cp}$) in Fig.\ref{fig3}d. The mode resides within the 1H layer and is localized to the bottom of the step edge of the covering 1T layer. This suggests that the 1T encapsulation alters the topological class of the superconducting state in the crater surroundings, thus inducing a topological transition at the crater's boundary.

The last important spectroscopic observation on the nature of the edge mode we obtain by following its character along the crater's varying boundary. In Fig.\ref{fig3}e we spatially map the ZBC along the crater's boundary. We find that the mode along the zigzag 1T edge appears narrower than the one bound to the other edge, which seems aligned with the 1T CDW pattern (marked by blue and red arrows, respectively). We quantify their extent by plotting their respective spatial profiles in Fig.\ref{fig3}f. While both seem to be exponentially localized at the crater boundary, the localization length of the mode running along the zigzag edge ($\sim$13 nm) is about 30$\%$ shorter than the one running along the other edge ($\sim$18 nm). The latter is similar to the localization length we found along the 1H-TaS$_2$ step edge terminations that agree also with the coherence length extracted from vortex core profiles. The observed anisotropy of the localization lengths suggests an anisotropic gap structure, commonly signifying nodal superconductivity. Albeit, the isotropic nature of the imaged vortex cores suggest a mild nodal structure, as nodal superconducting gaps are typically reflected in the vortex core structure. 

Before resolving their physical origin, we first summarize the spectroscopic observations we obtained on 4Hb-TaS$_2$ below its superconducting transition: (1) Exposed 1H-TaS$_2$ layers host exponentially localized edge modes on their crystallographic boundaries that have no trace within the normal state. (2) A similar mode is found on superconducting 1H-TaS$_2$ layers, at the boundaries between 1T-encapsulated and exposed regions of the same layer. (3) The edge mode displays anisotropic localization length, while vortices show isotropic core profile. (4) Vortices exhibit ZBC peaks at their cores. (5) A spatially uniform residual DOS is observed within the superconducting gap throughout the 1H-TaS$_2$ surface. These findings provide evidence for the existence of a topological superconducting state in the 1H-TaS$_2$ whose topological classification is rooted either in the interaction between the 1H-TaS$_2$ and adjacent 1T layers or by the symmetry breaking induced by its exposure on the surface. Several theoretical models have considered various realizations of topological superconducting states in TMDs \cite{Yuan2014,Zhou2016,Hsu2017,He2018TSC, Fischer2018,Hsu2020,Kanasugi2020,Shaffer2020}. However, they do not seem to capture all aspects of our spectroscopic observations. Here, we propose a simple model for topological superconductivity in which on-site cross-orbital pairing is favored over a standard $s$-wave like paring. The model requires mirror symmetry breaking, and yields a nodal point Weyl-like topological superconducting state. The detected anisotropic edge modes are thus analogous to Fermi arc boundary modes in topological nodal point semimetals. As we show below all aspects of our data are captured surprisingly well within this model.

\begin{figure}
\includegraphics[scale=1.25]{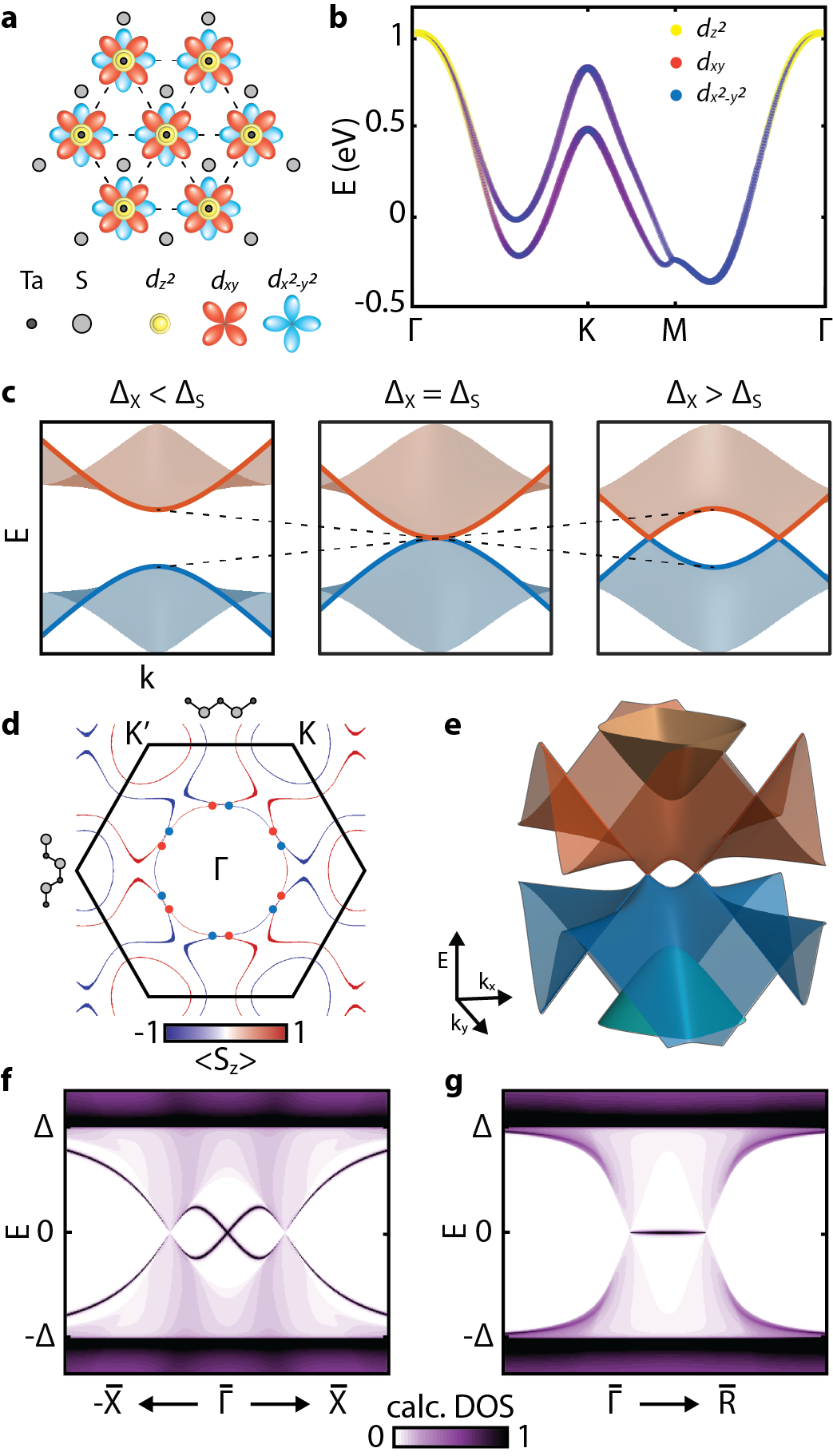} 
\centering
\caption{}     \label{fig4}
\end{figure}

We thus turn to describe our model of the inter-orbital pairing state in 4Hb-TaS$_2$ (more details in the supplementary section 11). The dominant orbital contribution at the Fermi level in a TMD monolayer comes mostly from the 4$d$ transition metal orbitals rather than the chalcogen $p$ orbitals \cite{Mockli2018}. It is, therefore, possible to construct a low-energy three-orbital tight-binding model, taking into account a minimal basis set of orbitals including $d_{z^2}$, $d_{xy}$ and $d_{x^2-y^2}$ (Fig.\ref{fig4}a, yellow, red and blue, respectively). The resulting band structure, including spin orbit coupling, plotted along high symmetry lines in Fig.\ref{fig4}b, has states at the Fermi energy with mixed orbital texture, depicted by colors. Typically, $s$-wave Cooper-pairing, $\Delta_S$, is assumed to be dominant among states of same orbital texture that necessitates pairing of singlet spin states. However, the system symmetry also permits crossed-orbital pairing, $\Delta_X$, characterized by an odd superposition of orbital states with even triplet spin states \cite{Gao2010,Wang2012,Fukaya2018}. The intra- and crossed-orbital pairing channels compete, and when the latter becomes dominant nodal point superconductivity is induced, as depicted in Fig.\ref{fig4}c. We thus attribute the residual DOS that offsets the BCS-type spectrum (Fig.\ref{fig1}f) to in-gap nodal states. We do not resolve here the microscopic mechanisms that favor crossed-orbital pairing. Nevertheless, electron interactions may very well have a substantial role in enhancing it, as crossed orbitals will minimize the on-site charge overlap (see Fig.\ref{fig4}a). Interactions are strong in TMD compounds and their effects are abundant, including the CDW order in TaS$_2$ that is thought to give rise to a Mott insulating state in bulk 1T. This is apparent on 1T layers of the 4Hb polytype, and is also imprinted in the local DOS of the adjacent 1H layers.

The transition from a fully gapped $s$-wave pairing to nodal-point triplet pairing constitutes a topological phase transition. Our theoretical calculation for the 4Hb-TaS$_2$ system shows that the nodal point superconducting state is characterized by 6 pairs of non-degenerate Dirac cones across the 2D Brillouin zone, as shown in Fig.\ref{fig4}d. Those in-gap Dirac cones are singly degenerate and posses well defined chirality \cite{Wang2012,Schnyder2015,Fukaya2018} of opposite signs among time-reversed partners, akin to Weyl cones in a topological semimetal. The separation of the Dirac nodes, visualized in Fig.\ref{fig4}e, is given by $b=\pm\sqrt{\Delta_X-\Delta_S}/v_F$. The vanishingly small in-gap DOS within an otherwise gapped spectrum does not degrade much the isotropic structure of vortex cores (Fig.\ref{fig1}g). However, when the bulk chiral bands are projected to an edge they induce a corresponding topological edge mode, akin to Fermi arc modes on the boundaries of a Weyl semimetal. In the edge-projected 1D Brillouin zone the edge mode spans all momenta between the projected pairs of Dirac nodes with opposite chirality, as shown in Fig.\ref{fig4}f and g for a zigzag edge and arbitrarily oriented one, respectively (see also Fig.S12). These edge modes are the ones detected in our experiment at the 1D boundary of the 1H terminated step-edge and at the boundary of the 1H terminated crater region. 

Nevertheless, the existence of an edge mode at the boundary between exposed and 1T-encapsulated 1H layers signifies a topological phase transition induced by the interaction with the adjacent 1T layers. According to our model, the origin of this transition lies in the mirror symmetry-breaking nature of the inter-orbital pairing term, $\Delta_X$. Accordingly, the 1T-encapsulated superconducting 1H layer, representative of bulk behavior, exhibits a rather symmetric electronic environment (see Fig.\ref{fig1}a), whereas exposed 1H layers are mirror symmetry broken. This is greatly amplified by the strong interaction between 1T and 1H layers. Our ab initio calculations find a substantial charge (and spin) transfer from the CDW localized state on the 1T layer to the adjacent 1H layer where it remains rather localized (See Fig.S11) as indeed captured in our measurements (Fig.1d). The charge transfer induces a local dipolar field between the two layers. Its effect on the 1H layer is balanced off in the symmetrically encapsulated case, but sustains in the exposed 1H layer that has only a 1T layer beneath it.

Moreover, the momentum space distribution of in-gap nodal points, that induces the edge mode, also entails its anisotropic nature. Both the mode dispersion and its localization length depend on the crystallographic direction to which it is projected. Projection onto the zigzag edge along $\Gamma-M$ yields an electron-hole symmetric pair of dispersing modes among the doubly degenerate projected Dirac nodes (Fig.\ref{fig4}f). Its crossing point at $\bar{\Gamma}$ is protected by a combined mirror and time-reversal operation, $M_xT$, and its localization length on the zigzag edge is the shortest (Fig.S12). As the projection deviates from $\Gamma-M$ onto arbitrary directions we find a singly degenerate non-dispersing mode stretching between the edge projected nodes (Fig.\ref{fig4}g). While our energy resolution does not allow to discern the varying dispersion of the edge modes along various crystallographic directions, its effect on the localization length agrees with the trend our model finds. In both, the edge mode localization is shortest along the zigzag direction and increases away from it. The overall order of magnitude discrepancy in the values the model finds can be partially attributed to the renormalized Fermi velocity \cite{Wijayaratne2017} that the model neglects. 

Finaly, we note that our measurements cannot determine whether the topological superconductivity observed on the surface is time-reversal symmetric or not. Our ab initio calculations find that the charge transfer from the 1T to the 1H layer is accompanied also by transfer of magnetic moment, which may very well give rise to a magnetically ordered states, once the magnetic moments are embedded within the metallic 1H layer (see supplementary section 10). Indeed, enhanced muon relaxation was reported to onset concurrently below the superconducting transition \cite{Ribak2019}. Our theoretical calculations show that adding a time-reversal symmetry breaking term does not eliminate the edge modes from any of the edge directions but only slightly affects their dispersion in a way that cannot be distinguished by our spectroscopic measurements (see supplementary for more details). Spontaneous time-reversal symmetry breaking may support other origins for the observed topological superconducting state \cite{Kezilebieke2020} and should thus be further investigated.

In summary, we provide direct evidence for topological superconductivity on the surface of 4Hb-TaS$_2$. Our measurements reveal in-gap edge modes at the boundaries of 1H-terminated step-edges and craters along with an anisotropic localization length of the edge modes. These observations are naturally captured within a simple model for nodal point topological superconductivity where inter orbital paring is dominating over the standard $s$-wave pairing. This state is favored on 1H terminated surfaces due to 1H-1T layer coupling which leads to mirror symmetry breaking across the 1H layers. While more likely to occur on the surface, it is not ruled out that this state would form also in bulk 1H layers while inducing spontaneous mirror symmetry breaking within the bulk layers. 

\section{Methods}
\subsection{Sample growth}
High-quality single crystals of 4Hb-TaS$_2$ were prepared using the chemical vapor transport method. The appropriate amounts of Ta and S were ground and mixed with a small amount of Se (1\% of the S amount). The powder was sealed in a Quartz ampoule, and a small amount of iodine was added as a transport agent. The ampoule was placed in a three-zone furnace such that the powder is in the hot zone. After 30 days, single crystals with a typical size of 5 mm$\times$5 mm$\times$0.1 mm grew in the cold zone of the furnace.

\subsection{STM measurements}
The 4Hb-TaS$_2$ single crystals were cleaved in the scanning tunnelling microscope load lock at ultra-high vacuum conditions and at room temperature. The STM measurements were performed using commercial Pt-Ir tips. The Pt-Ir tips were characterized on a freshly prepared Cu(111) single crystal. This process ensured a stable tip with reproducible results. All $dI/dV$ measurements were taken using standard lock-in techniques.

\subsection{Density functional theory calculations}
We performed the density-functional theory (DFT) calculation follows in the framework of the generalized gradient approximation \cite{Perdew1996}) with the Vienna \textit{ab-intio} package \cite{Kresse1999}. We employed the PBE-D2 method to describe vdW interaction \cite{Grimme2006}. We included spin-orbit coupling (SOC) interaction in all calculations. The 4Hb structure with $\sqrt{13}\times\sqrt{13}$ supercell was used for the CDW reconstructed structure. Full geometry optimization was performed until the Hellmann-Feynmann force acting on each atom became smaller than 0.01 eV/\AA.

\clearpage

\clearpage
\textbf{Fig. 1 | Charge density wave and zero-bias peak in the vortex.} \textbf{a,} Crystal structure of 4Hb-TaS$_2$. \textbf{b,} Topography of a cleaved surface showing both the 1H and the 1T terminations along with charge density wave (CDW). \textbf{c,d,} Zoomed in topography of the 1T and the 1H termination, respectively (upper panel) and their corresponding Fourier transform (lower panel). The Ta and S atoms are overlaid on the topography along with the CDW unit cell (green rhombus). The Fourier transform clearly shows the atomic Bragg peak (yellow), the 1T CDW peak (green) and the 1H CDW peak (red). \textbf{e,} Spatially averaged dI/dV measured on the 1T and the 1H terminations at 4.2K. \textbf{f,} dI/dV profile (blue) measured on the 1H termination at T=0.38K clearly showing the superconducting (SC) gap. The normal state is recovered above $T_c$ (orange) and $H_{c2}$(purple). The SC gap fits well to BCS spectrum with finite offset (solid black). \textbf{g,} Spatial mapping of dI/dV(0) on the 1H termination with a finite out-of-plane magnetic field (B$_z$=100 mT) shows a well ordered Abrikosov lattice of vortices in a dilute background of speckles. \textbf{h,} dI/dV map across a vortex marked by the dashed line in \textbf{g} shows enhanced zero-bias density of states (DOS) at the core. \textbf{i,} dI/dV profiles at the vortex core (red) and far from it (blue). The corresponding locations are marked by red and blue arrows in \textbf{h}. \textbf{j} Zoom in of the region marked by the shaded box in \textbf{i}, clearly showing a zero-bias peak fit to a single Gaussian profile after instrumental broadening.

\clearpage

\clearpage
\textbf{Fig. 2 | Spatial mapping of the dispersing edge mode at the 1H step edge.} \textbf{a,} Topography of the 1H terrace terminating at a step edge (dashed black line) of a single 1H layer height, thereby revealing the 1T termination below, is shown in the upper panel. The lower panel shows the height profile of the step edge. \textbf{b,} Averaged dI/dV spectra measured close to the step edge (pale orange) and far from the step edge (pale blue), as marked by the red and blue filled circles in \textbf{a}, showing a distinct increase in the in-gap density of states (DOS) near the step edge. \textbf{c,} Spatially and energy resolved dI/dV measured along the grey line in \textbf{a} across the step edge (dashed line). \textbf{d,} (Upper panel) The spatial dependence of the position of the coherence peaks (E$_{cp}$) as extracted from \textbf{c}. (Lower panel) The zero bias dI/dV profile originating from the step edge (X=0 nm) is shown in filled blue circles. The dI/dV shows an exponential increase on approaching the step edge, indicating an edge mode localized to the step edge within the SC gap. An exponential fit to the dI/dV profile, shown with a black dashed line, results in a localization length of 18(1) nm, similar to the SC coherence length. \textbf{e,} Spatially resolved zero bias dI/dV map overlaid on topography, in the absence of an magnetic field. The edge mode appears all along the rough step edge (dashed line) of 1H terrace, indicating its robust spatial extent. \textbf{f,} The localization length of the edge mode ($l$) was extracted from the exponential fits, as shown in \textbf{d}, for several bias energies within the SC gap, is plotted in blue squares. A standard deviation of $\sigma$=1, is plotted in dotted lines. Scanning parameters: $V_{set}$ = 1.5 mV, $I_{set}$ = 50 pA, $V_{ac}$ = 50 $\mu$V, f = 773 Hz. a.u., arbitrary unit

\clearpage

\clearpage
\textbf{Fig. 3 | Anisotropic edge mode at the 1T step edge imprinted on 1H.} \textbf{a,} Topography of a large clean crater in the 1T termination of 4Hb-TaS$_2$, exposing the lower 1H layer. The black arrows show the 1T CDW directions. The height profile of the single layer step edges and terraces along the white dotted line is shown in the lower panel. \textbf{b,} Spatially resolved dI/dV(0) map measured with an out-of-plane field of 50 mT, in region 'b' in \textbf{a}, overlaid on top of the topography, showing a partially exposed vortex on the 1H termination. Inset shows a large scale zoomed out map of the vortex lattice in the same region. \textbf{c,} dI/dV map measured along the dashed line 'c' marked in \textbf{a}. \textbf{d,} The spatial dependence of the position of the coherence peaks (E$_{cp}$), extracted from \textbf{c}. \textbf{e,} Spatially resolved dI/dV(0) map, measured in region 'f' in \textbf{a}, showing a clear continuous edge mode localized to the step edges of different direction. The 1T step edge is marked by the black dashed line in \textbf{b} and \textbf{e}. \textbf{f,} The profile of zero bias conductance, starting from the step edge was extracted perpendicular to edge 1 and edge 2, as indicated by the blue and red arrows in \textbf{e}. The ZBC profile from the step edge along a direction similar to edge 1, marked as c in \textbf{a}, is shown in black. An exponential fit to the ZBC perpendicular to edge 1 yields a localization length, $l$ = 12(1) nm and 13(1) nm while along edge 2, $l$ = 18(2) nm. Scanning parameters for \textbf{b} and \textbf{e}: $V_{set}$ = 2 mV, $I_{set}$ = 100 pA, $V_{ac}$ = 100 $\mu$V, f = 773 Hz. a.u., arbitrary unit.

\clearpage

\clearpage
\textbf{Fig. 4 | Inter-orbital pairing and topological nodal superconductivity.} \textbf{a,} Schematic illustration of the various $d$-orbitals of Ta decorating the lattice of 1H-TaS$_2$. \textbf{b,} Calculated band structure of 1H-TaS$_2$ with different orbital contribution encoded in color. \textbf{c,} Schematic illustration of a topological phase transition with increasing $\Delta_X/\Delta_S$, where $\Delta_X$ and $\Delta_S$ are the inter and intra-orbital pairing amplitude. \textbf{d,} Calculated Fermi surface of 1H-TaS$_2$. The chirality of the topological nodal points is shown in red and blue circles. \textbf{e,} A schematic showing a pair of topological nodal points for $\Delta_X>\Delta_S$. \textbf{f,g,} Calculated density of states (DOS) along the zig-zag and arbitrary edge, showing dispersing and flat Majorana edge mode, respectively. 


%





\clearpage
\bibliography{Mendeley_1HTaS2}

\section{Data availability}
The data that support the plots within this paper and other findings of this study are available from the corresponding authors upon reasonable request.

\section{Code availability}
The codes used in theoretical simulations and calculations are available from the corresponding authors on reasonable request.

\section{Acknowledgements}
N.A., H.B., and B.Y acknowledge the German–Israeli Foundation for Scientific Research and Development (GIF grant no. I-1364-303.7/2016). H.B. and N.A. acknowledge the European Research Council (ERC, project no. TOPO NW), B.Y. acknowledges financial support by the Willner Family Leadership Institute for the Weizmann Institute of Science, the Benoziyo Endowment Fund for the Advancement of Science, the Ruth and Herman Albert Scholars Program for New Scientists, and the Israel Science Foundation (ISF 1251/19). G.A.F. gratefully acknowledges partial support from the National Science Foundation through NSF Grant no. DMR-1720595, and DMR-1949701. Y.O. acknowledges partial support through the ERC under the European Union’s Horizon 2020 research and innovation programme (grant agreement LEGOTOP No 788715), the ISF Quantum Science and Technology (2074/19), the BSF and NSF (2018643), and the CRC/Transregio 183. A.K. acknowledges the Israel Science Foundation (ISF 320/17).

\section{Author contributions}
A.K.N., A.S. and Y.R. acquired and analyzed the data. N.A. and H.B. conceived the experiments. J.K. and B.Y. calculated the ab initio model. G.M., G.A.F, B.Y., and Y.O. calculated the theoretical model. A.K. grew the material. A.K.N., N.A., and H.B. wrote the manuscript with substantial contributions from all authors.

\section{Competing interests}
The authors declare no competing interests.

\section{Supplementary information}
The supplementary information file contains sections 1-11 and figures S1-S14.

\end{document}